\documentstyle[epsfig]{aipproc}

\begin{document}
\title{Scattering  of plane-wave atomic vibrations  in disordered 
structures}

\author{S.N.Taraskin and S.R.Elliott}
\address{Department of Chemistry, 
University of Cambridge, 
Lensfield Road,
 Cambridge CB2 1EW, UK}

\maketitle 
\vskip 20pt

\centerline{(to be published in the Proceedings 
of the summer school (Cargese, 1999):} 
\centerline{ {\it Physics of Glasses: 
Structure and Dynamics}, Eds. P. Jund and R. Jullien)}

\begin{abstract} 
A theoretical analysis of the scattering 
of plane-wave atomic excitations in disordered 
solids has been made
 in terms of the spectral densities. 
Hybridization between transverse and longitudinal waves 
of approximately the same frequency is demonstrated. 
The analytic results agree well with the results obtained from 
computer  simulation for a toy linear zig-zag chain model 
and a model of vitreous silica constructed by molecular dynamics. 
\end{abstract}

\section{Introduction}
\label{S1}
Propagation of classical plane waves in random 
scattering media has attracted a lot of 
theoretical and experimental attention 
in recent years 
\cite{PingSheng_book1,PingSheng_book2}. 
The phenomena of wide interest include  
weak localization \cite{PingSheng_book1,PingSheng_book2}, 
 Anderson localization \cite{Mott_Davis_book},  
phonon localization \cite{John,Aharony,Alexander}  and related 
behaviour of  plane waves in the Ioffe-Regel crossover  
region \cite{IR} separating 
weakly and strongly scattered waves 
\cite{PingSheng_1994,Anglaret}. 

Vibrational plane waves propagate well in disordered structures in the 
long-wavelength limit, 
$ka \ll 1$, with $k$ the wavevector and $a$  a measure of 
the microscopic scale of the structure (being of the  order 
of interatomic distances), when the atomic structure 
behaves as an elastic continuum and  disorder on the microscopic 
level is not of great importance. 
The situation is changed with decreasing  wavelength and, on 
the microscopic scale, disorder becomes 
important and the wavevector is no longer  a good quantum value 
\cite{Grest,Hafner,Schober_Liard,Gurevich_93,Mazzacurati_EPL,Sampoli_98,EPL,Madden}. 

In the investigation of the propagation of 
the plane-wave vibrational excitations in
 disordered structures,  
different decay channels have been suggested to explain their 
attenuation: 
(i) disorder-induced channels \cite{Allen1,Allen,Fabian,DelAnna,Feldman_1998},  
(ii) anharmonic channels \cite{Fabian} and 
(iii)  channels involving  two-level systems 
 \cite{Wischnewski_98,PingSheng_1991,SM_phonons,Burin}. 
The anharmonic channels are strongly enhanced 
with increasing temperature, particularly at temperatures 
comparable with the glass-transition temperature, $T_{\rm g} \sim 10^3$K. 
In contrast,  scattering by  two-level systems 
can be  important at low temperatures \cite{SM_phonons}, 
$T\ll T_{\rm TLS}\sim 10 - 100$K \cite{PingSheng_1991,Vacher}. 
In the intermediate temperature range, $T_{\rm TLS} 
\le T \ll T_{\rm g}$, which is considered below,  the scattering 
processes involving  
two-level systems are suppressed and the 
atomic dynamics are usually harmonic  \cite{DelAnna,Kob_JNCS}, 
meaning that  disorder-induced channels play 
the most important role in the decay mechanism 
of plane-wave excitations. 

If  the harmonic approximation is valid for describing the atomic 
dynamics,  then a normal-mode analysis can be used for the problem 
 under consideration. 
The normal modes can be found  either analytically 
or numerically. 
A general theory of atomic vibrations in disordered structures, 
which in principle should 
result in normal modes,  
has been  mainly developed for particular simple 
model structures  
\cite{Allen1,Maradudin,part1,part2,Klinger_review,Karpov_review,Schirmacher}, 
which can hardly describe quantitatively the 
situation in real structures. 
Therefore, 
a  numerical approach could be 
very useful in the calculation of normal modes, e.g. 
by direct diagonalization of the dynamical matrix which can be available, e.g. 
from molecular dynamics simulations. 

The main questions we address in  this paper are: 
 how are plane-wave vibrational excitations scattered by  
disorder and what are the characteristics of the final 
state after scattering? 

Our approach to the problem 
is based on combining analytical and numerical techniques. 
First, we create a few  structural models of a disordered atomic material: 
(i)   realistic models of v-SiO$_2$  
using molecular dynamics and  
(ii) toy models of a linear zig-zag chain. 
Then all eigenmodes and eigenfrequencies (in the harmonic approximation) 
are found numerically. 
These characteristics fully determine the dynamical response 
of the system (and final state at $t\to \infty$) 
 to any external excitations, including the plane-wave 
excitations  of present interest. 
The final state after scattering was investigated then 
 in momentum space (analytically and numerically). 
\section{Formalism}
\label{S2}
The time evolution of any vibrational excitation is fully 
determined in the  harmonic approximation by the eigenmodes 
and eigenfrequencies of the system. 
Indeed, the initial excitation can be expanded in 
the eigenmodes, the time dependence of which is known. 
The coefficients in such an expansion are defined by the shape 
of the initial vibrational excitation. 
Here we consider only plane-wave external initial 
excitations, mainly because exactly such excitations 
are generated in a system by inelastic neutron, light and electron 
scattering \cite{SRE_book}.

In amorphous  materials, because of disorder, the eigenmodes are not plane 
waves even in the long-wavelength limit. 
Therefore, an initial plane wave, when expanded over eigenmodes, 
contains different eigenmodes characterized by different weights in 
this expansion. 
The eigenmodes participating in the expansion evolve differently with time, 
  so that the propagating  excitation becomes 
 different in shape  compared with the initial one. 
If we expand the vibrational state in plane waves after a 
certain time, then this expansion contains not only 
the initial plane-wave component but also  other 
plane waves characterized by different wavevectors. 
This means that the initial plane wave is scattered by the  
structure into a different final state. 
Our aim here is to study the properties of the final state after 
decay, for different wavevectors of an initial plane wave. 

Let us consider an external wave  excitation introduced 
in the system, 
${\bf u}(t)$, 
which at the initial moment of time is an ideal plane wave, 
${\bf w}_{\bf k}$,
characterized by the wavevector ${\bf k}$, 
 unit polarization vector $ {\bf {\hat n}}$ and initial phase $\phi_0$:  
\begin{equation}
{\bf u}(t=0)= {\bf w_k} \equiv 
A{\bf {\hat n}} \cos[ {\bf k\cdot r}+\phi_0] 
\ ,  
\label{e3_1}
\end{equation}
where ${\bf u}$ is a $3N$-dimensional displacement 
vector, the $i$-th component of which describes 
the displacement of atom $i$ from its equilibrium position 
at ${\bf r_i}$, $A$ is the normalization constant defined 
below and the wavevector index ${\bf k}$ includes 
also the polarization index ${\bf {\hat n}}$.  
In our analytical treatment, we assume that 
eigenmodes and eigenfrequencies are known, e.g. 
from numerical simulations. 
The initial displacement vector, Eq.~(\ref{e3_1}), each atomic 
component of which is multiplied by the mass 
factor $m_i=M_iN/\sum_i M_i$ ($M_i$ stands for the mass 
of atom $i$) can be expanded in eigenmodes as: 
\begin{equation}
{\bf u}(0) =
\sum_{j=1}^{3N} 
{\overline \alpha^j_{\rm {\bf k}} } 
{ {\bf  e^j }/ \sqrt{m} } \ , 
\label{e3_2}
\end{equation}
where the symbolic script ${\bf  e^j }/\sqrt{m}$ means 
that each $i$-th component of vector ${\bf e^j}$ is divided 
by the factor $\sqrt{m_i}$.  
The coefficients $ {\overline \alpha^j_{\rm {\bf k}} } $ 
in expansion (\ref{e3_2}), the squares of which are the spectral densities 
of  the system,  
are defined by the following equation: 
\begin{equation}
{\overline \alpha^j_{\rm {\bf k}} } = 
\langle {\bf e}^j \cdot \sqrt{m} {\bf u}_{\rm {\bf k}}\rangle \equiv 
\sum_{i=1}^N  \sqrt{m_i} {\bf e}^j_i \cdot {\bf w}_{{\rm {\bf k}},i}
\ . 
\label{e3_3}
\end{equation}
The spectral-density coefficients, Eq.\  (\ref{e3_3}), fully determine 
the dynamical response of the system to plane-wave 
excitation. 
Indeed, at any moment of time $t$, the displacement 
vector of the propagating excitation can be 
represented via eigenmodes developing in time as: 
\begin{equation}
{\bf u}(t)=\sum_{1}^{3N} 
{\overline \alpha^j_{\rm {\bf k}} } 
{ {\bf e}^j \over \sqrt{m} } 
\cos\omega_jt 
\ . 
\label{e3_4}
\end{equation}
For the sake of simplicity and without loss of generality 
(as  shown below), we 
consider the initial excitation to be a standing wave, 
i.e. $ \dot{\bf u}(0)=0$, leading to the 
absence of terms proportional to $ \sin\omega_jt$ 
in expression (\ref{e3_4}).  
It is convenient for the initial vector 
$\sqrt{m} {\bf u_k}(0) $ to be normalized to 
unity, so that 
\begin{equation}
\sum_{1}^{3N} 
|{\overline \alpha^j_{\rm {\bf k}} } |^2 
= 1  
\ ,  
\label{e3_5}
\end{equation}
and the normalization constant in Eq.~(\ref{e3_1}) is 
$A^2=\left[\sum_i m_i |{\bf u_k}(0)|^2\right]^{-1}$.

An ideal initial plane wave (\ref{e3_1}) scatters 
with time to different plane waves. 
In order to calculate the weights of  different plane-wave 
components in the propagating excitation, we 
expand  the displacement vector 
$ {\bf u}(t)$ in plane waves: 
\begin{equation}
{\bf u}(t)= \sum_{{\rm {\bf k'}}}
{\bf u}_{{\rm {\bf k'}}}(t) 
\ ,
\label{e3_6}
\end{equation}
where the sum is taken over all wavevectors  ${\bf k'}$ 
(allowed by the periodic simulation box in the 
case of a finite model) 
and all polarizations (two transverse and one longitudinal for each 
wavevector), and the waves 
$ {\bf u}_{{\rm {\bf k'}}}(t) $ are defined as: 
\begin{equation} 
{\bf u}_{\rm {\bf k'}}(t) 
= a_{\rm {\bf k'}}(t) A  {\bf {\hat n}'} 
 \cos{\left({\bf k' \cdot r} + \phi_{\rm {\bf k'}}(t)\right)} 
\ .  
\label{e3_7}
\end{equation}
The same normalization  as in Eq.~(\ref{e3_1}) is used here. 
In order to find the time dependence of the 
amplitude $a_{\rm {\bf k'}}(t)$ and phase 
$\phi_{\rm {\bf k'}}(t)$, it is convenient to 
rewrite Eq.~(\ref{e3_7}) in the following form: 
\begin{equation} 
{\bf u}_{\rm {\bf k'}}(t)
 = 
 a_{\rm {\bf k'},c}(t) {\bf w}_{\rm {\bf k'},c} + 
a_{\rm{\bf k'} ,s}(t){\bf w}_{\rm {\bf k'},s}
\ ,
\label{e3_8}
\end{equation} 
where 
\begin{equation}
 {\bf w}_{\rm {\bf k'},c} = A {\bf {\hat n'}} \cos{\bf k'\cdot r}\ \ \ 
\mbox{and}  \ \ \ 
 {\bf w}_{\rm {\bf k'},s} = A {\bf{\hat n'}} \sin{\bf k'\cdot r}
\ , 
\label{e3_9}
\end{equation}
so that 
\begin{eqnarray}
a_{\rm {\bf k'}}(t)&=&( a^2_{\rm {\bf k'},c}(t) + 
a^2_{\rm {\bf k'},s}(t))^{1/2} \ , 
\label{e3_10}
\\ 
\phi_{\rm {\bf k'}}(t)&=&  
\mbox{Arctan}[a_{\rm {\bf k'},s}(t)/ a_{\rm {\bf k'},c}(t)]
\ .
\label{e3_11}
\end{eqnarray} 
The coefficients $ a_{{\bf k'},{\rm c(s)}}(t) $  before  
the $\cos$- ($\sin$-) like components in Eq.~(\ref{e3_8}) 
can be found by multiplying both sides of this equation 
by ${\bf w}_{ {\bf k'},{\rm c(s)}}$ and using Eq.~(\ref{e3_3}), so that 
\begin{equation}
 a_{\rm k',c(s)}(t) =\sum_j 
{\overline \alpha}^j_{\rm {\bf k}}
{\underline \alpha}^j_{{\rm {\bf k'}},{\rm c(s)}} \cos\omega_jt / 
\langle {\bf w}_{\rm {\bf k'},c(s)}^2 \rangle
\ .  
\label{e3_12}
\end{equation}
where 
\begin{equation}
\langle {\bf w}_{\rm {\bf k'},c(s)}^2 \rangle \equiv 
\sum_{i=1}^{N}  |({\bf w}_{{\rm {\bf k'},c(s)}})_i|^2 
\ .  
\label{e3_12a}
\end{equation}
The spectral-density coefficients 
$ {\underline \alpha}^j_{{\rm {\bf k'}},{\rm c(s)}}$ 
for plane waves of $\cos$- and $\sin$-like type 
are defined by the following equation: 
\begin{equation}
{\underline \alpha}^j_{{\rm {\bf k}},{\rm c(s)}} = 
\langle 
{\bf e}^j \cdot  {1 \over \sqrt{m}} {\bf w}_{\rm {\bf k},c(s)}
\rangle  \equiv 
\sum_{i=1}^N {1 \over \sqrt{m_i}} {\bf e}^j_i \cdot ({\bf w}_{{\rm {\bf k},c(s)}})_i
\ . 
\label{e3_13}
\end{equation}
The spectral-density coefficients 
$ {\underline \alpha}^j_{{\rm {\bf k}},{\rm c(s)}}$ 
in a multicomponent system,  
in contrast to 
${\overline \alpha}^j_{\rm {\bf k}}$, instead of being normalized to unity 
are  normalized by the following value: 
\begin{equation}
\sum_{j=1}^{3N} 
|{\underline \alpha^j_{\rm {\bf k}, c(s)} } |^2 
= 
{ \sum_i (m_i)^{-1} |{\bf w}_{\rm {\bf k},c(s)}|^2 \over 
\sum_i m_i |{\bf w}_{\rm {\bf k},c(s)}|^2 } 
\ ,  
\label{e3_14}
\end{equation}
having the value  $\simeq 0.7$ in the case of vitreous silica.

 Eqs.\ (\ref{e3_10}) - (\ref{e3_14}) fully 
determine the time evolution of different 
${\bf k'}$ plane-wave components in the propagating 
vibrational excitation via the spectral densities and 
the vibrational spectrum itself. 
Another useful characteristic often used to characterize 
the decay of an initial external excitation is the time 
correlation function \cite{Hansen_book}, 
\begin{equation}
{ \langle {\bf u}(t)  
\cdot {\bf u}(0) \rangle  
\over 
\langle {\bf u}(0)  
\cdot {\bf u}(0) \rangle  }
\equiv 
{ \sum_i {\bf u}_i(t) {\bf u}_i(0) \over 
\sum_i {\bf w}_{{\rm {\bf k}},i} 
{\bf w}_{{\rm {\bf k}},i}   } 
= 
{ \sum {\overline \alpha}^j_{\rm {\bf k}}
{\underline \alpha}^j_{\rm {\bf k}}
\cos\omega_jt \over \langle {\bf w}^2_{\rm {\bf k}} 
\rangle } 
\ , 
\label{e3_15}
\end{equation}
where the spectral-density coefficient 
$ {\underline \alpha}^j_{\rm {\bf k}} $ is defined 
by Eq.~(\ref{e3_13}) with 
${\bf w}_{\rm {\bf k},c(s)}$ replaced by 
${\bf w}_{\rm {\bf k}}$, and is related to the spectral-density 
coefficient 
${\overline \alpha}^j_{\rm {\bf k}}$ according 
to the following equation: 
\begin{equation}
{\underline \alpha}^j_{\rm {\bf k}} = 
\sum_{j'} {\overline \alpha}^{j'}_{\rm {\bf k}} 
 \langle {\bf e}^j m^{-1}   
 {\bf e}^{j'}  \rangle  
\ . 
\label{e3_16}
\end{equation}
In the case of a one-component system, the spectral-density coefficients  
${\underline \alpha}^j_{\rm {\bf k}}$ and 
${\overline \alpha}^j_{\rm {\bf k}}$ are obviously 
identical.

The decay of the external plane-wave excitation can 
 also be characterized via the properties of the final 
state after decay, averaged over time as $t \to \infty$. 
An initial plane-wave excitation characterized 
by the wavevector ${\bf k}$ and polarization 
${\hat{\bf n}}$ is scattered to different plane-wave 
components characterized 
by the wavevectors ${\bf k'}$ and polarizations 
${\bf {\hat n}'}$. 
The distribution, 
$ \rho({\bf k'},{\bf {\hat n}'}|{\bf k},{\bf{\hat n}}) $, 
of the weights 
of different plane-wave components averaged over 
time in the final state, 
$ \overline{a^2_{\bf k',{\hat n}'}(t) } $, is of  particular 
interest. 
Such a distribution is defined by the following equation:  
\begin{equation}
\rho({\bf k'},{\bf {\hat n}'}|{\bf k},{\bf{\hat n}})
= 
\overline{a^2_{\bf k'}(t) }
= {1\over 2} \left\{   
{\sum_{j}|{\overline\alpha}^j_{\rm {\bf k}}|^2
|{\underline\alpha}^j_{{\rm {\bf k'}},{\rm s}}|^2 
\over (\langle {\bf w}_{\rm {\bf k'},s}^2 \rangle )^2 } 
+  
{\sum_{j}|{\overline\alpha}^j_{\rm {\bf k}}|^2
|{\underline\alpha}^j_{{\rm {\bf k'}},{\rm c}}|^2 
\over (\langle {\bf w}_{\rm {\bf k'},c}^2 \rangle )^2 }  
\right\}  
\ ,  
\label{e3_17}
\end{equation}
where, as in Eq.~(\ref{e3_1}) and subsequently, 
the wavevector index ${\bf k'}$ also includes 
the polarization index ${\bf{\hat n}'}$. 
 Eq.~(\ref{e3_17}) is obtained by averaging 
Eq.~(\ref{e3_10}) over time  and using Eq.~(\ref{e3_12}). 
Bearing in mind that 
$ \langle {\bf w}_{{\bf k'},{\rm s}}^2 \rangle \simeq 
\langle {\bf w}_{{\bf k'},{\rm c}}^2 \rangle \simeq
\langle {\bf w}_{\bf k'}^2 \rangle 
$ 
and that the sum 
$ |{\underline\alpha}^j_{{\bf k'},{\rm s}}|^2 + 
|{\underline\alpha}^j_{{\bf k'},{\rm c}}|^2 $ 
is independent of the phase of $\cos$- and $\sin$-like 
components defined by Eq.~(\ref{e3_9}),  
expression (\ref{e3_17}) can be transformed to: 
\begin{equation}
\rho({\bf k'},{\bf {\hat n}'}|{\bf k},{\bf{\hat n}}) = 
{\sum_{j}|{\overline\alpha}^j_{\bf k}|^2
|{\underline\alpha}^j_{\bf k'}|^2 
\over  (\langle {\bf w}_{\bf k'}^2 \rangle )^2 } 
\ . 
\label{e3_18}
\end{equation}
The distribution (\ref{e3_18}) averaged over all polarizations 
${\bf{\hat n}'}$ in the final state is
\begin{equation}
\rho_{\rm av}({\bf k'}|{\bf k},{\bf{\hat n}}) \simeq 
2\rho({\bf k'},{\bf{\hat n}'}_{\rm t}|{\bf k},{\bf{\hat n}}) +
\rho({\bf k'},{\bf{\hat n}'}_{\rm l}|{\bf k},{\bf{\hat n}}) 
\ , 
\label{e3_19}
\end{equation}
where the unit vector $ {\bf{\hat n}'}_{\rm t} $ 
stands for transverse polarization in the final state 
while $ {\bf{\hat n}'}_{\rm l} $  refers to  
 longitudinal polarization,  
and the  factor $2$ takes into account the  existence of two 
independent and, in glasses,  equivalent  transverse polarizations. 
Glasses are isotropic, and an averaging of 
Eqs.\ (\ref{e3_17})-(\ref{e3_19}) over the 
directions of both initial and final wavevectors 
(including averaging over transverse polarizations 
in the initial wave) can be made,  
resulting in:
\begin{equation}
\rho_{\rm av,t(l)}(k'|k) =
\langle \rho_{\rm av}({\bf k'}|{\bf k},{\bf{\hat n}}_{\rm t(l)}) 
\rangle_{\Omega_{{\bf k},{\bf k'}}} 
\ .  
\label{e3_20}
\end{equation}

If we are interested in the contribution of the  same 
plane-wave 
${\bf k}$-component as in the initial excitation, then 
the wavevector ${\bf k'}$ should be replaced 
by ${\bf k}$ in 
Eqs.\ (\ref{e3_17})-(\ref{e3_19}) and 
 averaging only over  ${\bf k}$-directions 
should be made in Eq.~(\ref{e3_20}). 
%
%
%
%

%
%
%
%
\section{Spectral densities}
\label{S3}
As  follows from Sec.\ \ref{S2}, the coefficients  
${\overline\alpha}^j_{\bf k}$, 
${\underline\alpha}^j_{\bf k}$ in the expansion 
of  different ${\bf k}$-plane-waves over 
the eigenmodes, i.e. projections of plane waves onto 
eigenvectors, and related spectral densities, 
$ |{\underline\alpha}^j_{\bf k}|^2 $,
$ |{\overline\alpha}^j_{\bf k}|^2 $ and 
$ {\overline\alpha}^j_{\bf k} 
{\underline\alpha}^j_{\bf k} $, 
fully determine the dynamical response of the system 
to the initial plane-wave excitation. 
We calculate below the spectral densities for two 
models of disordered structures: 
(i) a model 
of v-SiO$_2$ constructed by molecular dynamics and    
(ii) a disordered zig-zag chain.  
%
%
%
%
%
%
%
\subsection{Vitreous silica}
\label{S3_1}
The models of v-SiO$_2$ have been constructed by 
$NPT$-molecular-dynamics 
simulations, 
using the potential of van Beest \cite{VB}. 
The van Beest potential has been modified  
for small interatomic distances according to 
Guissani and Guillot \cite{Guissani}. 
At large interatomic distances, we have used a cutoff  
for short-range interactions, multiplying the 
modified van Beest potential by a Fermi-like 
step function. 
The step function is  characterized by the step position 
at $R_{\rm cut}=5.5 \AA$  and the step width 
$\delta R_{\rm cut}=0.5 \AA$ for all atomic species. 
The latter cutoff has been used to obtain a  
density of the glassy structure (at zero pressure), 
 of $2.38$g$/$cm$^3$, 
reasonably close to the experimental value of 
$2.2$g$/$cm$^3$ (see the discussion of the 
densification problem in Ref.~\cite{Guissani}). 
Note that a similar cutoff ($R_{cut}=5.0\AA$ and 
$\delta R_{cut}=0$) has been used by Vollmayr {\it et al.} 
\cite{Kob}.
 
  All glassy models have been created by quenching from 
the melt ($T=6000$K) 
 to  the well-relaxed glassy state 
($T \sim 10^{-4}$K) at an average quench rate  of 
$\sim 1$K$/$ps. 
No coordination defects have been found in the models. 
The fully dense dynamical matrices for the relaxed 
systems were diagonalized directly, resulting in  
eigenvectors $\{ {\bf e}^j \} $ and eigenvalues $(\omega_j)$, 
thus allowing us to perform a complete 
harmonic vibrational analysis. 
Structural characteristics and 
vibrational properties of the models are very similar to 
those described in Ref.~\cite{PRB2}.

%
%
%
%
%
\begin{figure}[b!] %
\centerline{\epsfig{file=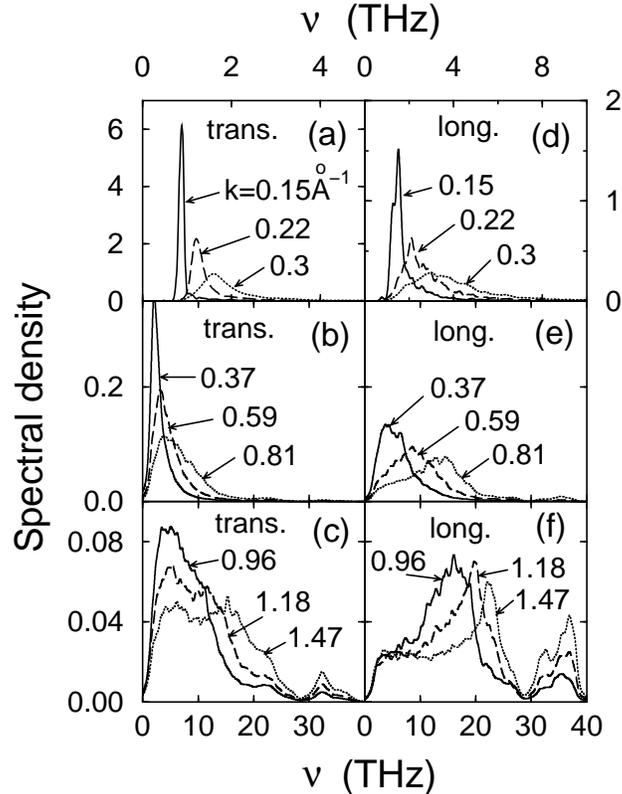,scale=0.5,angle=0}}
\vspace{1pt}
\caption{
 The spectral densities 
$|\overline{\alpha}^j_{\bf k}|^2$ 
for transverse ((a), (b) and (c)) and longitudinal 
((d),  (e) and (f)) 
initial polarizations for different 
magnitudes, $k$,  of the initial wavevector 
as shown in the figure.  
}
\label{f17}
\end{figure}

The models of v-SiO$_2$ were of two types: 
a cubic model  containing $N=1650$ 
atoms 
 and of box  length $L \simeq 28.4 \AA $, and  a bar 
configuration containing $N=1500$ atoms of size 
$85.6\AA\times 15.6\AA\times 15.6\AA$ 
(a bar-shaped model of B$_2$O$_3$ has 
been also used in Ref.\ \cite{Bermejo}). 
The bar-shaped models were constructed 
to allow access to much lower 
values of $k \ (\ge 0.07\AA^{-1})$ 
for modes propagating along the bar 
than can be obtained for the cubic models 
($k \ge 0.22\AA^{-1}$).

%
%
%
%
\begin{figure}[b!] %
\centerline{\epsfig{file=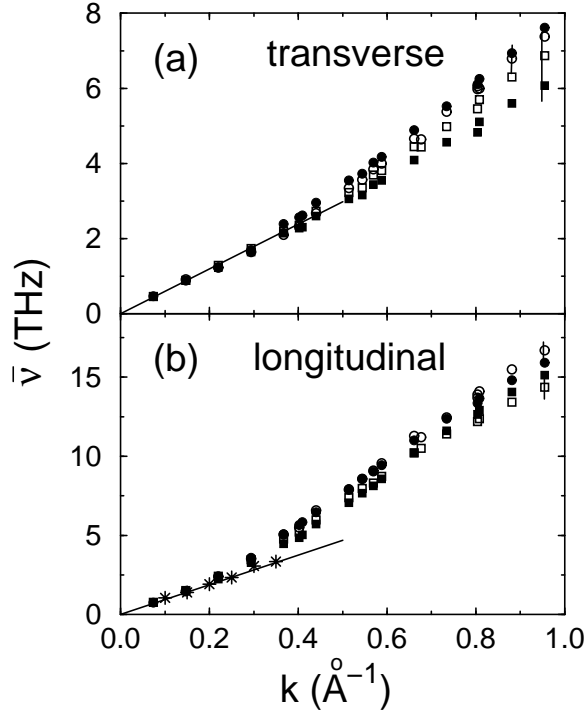,scale=0.5,angle=0}}
\vspace{10pt}
\caption{
The dispersion laws for transverse (a) and longitudinal 
(b) polarizations of the initial plane-wave excitation. 
The solid circles and squares were  
obtained from the fit of the spectral densities by 
Lorentzians and the DHO model, respectively.  
The stars in (b) correspond to IXS data by Benassi 
{\it et al } \protect\cite{Benassi}. 
The solid lines represent the long-wavelength 
limit characterized by 
the experimental sound velocities.  
}
\label{f18}
\end{figure}

Silica is a multicomponent system so that the 
three spectral densities differ from each other 
due to  different contributions 
in the coefficients  ${\overline\alpha}^j_{\bf k}$ and  
${\underline\alpha}^j_{\bf k}$ of the mass factor 
(see Eqs.\ (\ref{e3_3}) and (\ref{e3_13})).
In the case of vitreous silica, 
 the masses of the atomic species are quite comparable 
and the mass factor is of the  order of  unity, 
so that the different types of the spectral 
densities  differ only slightly from each other and 
the following approximate relationships 
can be used: 
$ 
|{\underline\alpha}^j_{\bf k}|^2 \simeq  A_1 
|{\overline\alpha}^j_{\bf k}|^2 
$
and
$ 
{\overline\alpha}^j_{\bf k} 
{\underline\alpha}^j_{\bf k} \simeq 
A_2  
|{\overline\alpha}^j_{\bf k}|^2 
$, 
with the normalization constants for the corresponding 
spectral densities being 
$ 
A_1 
= 
 \sum_j |{\underline\alpha}^j_{\bf k}|^2 
$ 
and 
$ 
A_2 
= 
\sum_j {\overline\alpha}^j_{\bf k} 
{\underline\alpha}^j_{\bf k} 
$, 
which,  in the case of vitreous silica, give  
$ A_1 \simeq 0.7$ and 
$ A_2 \simeq 0.8$.

The shape of the spectral density depends on the characteristics 
of the plane wave (wavevector and polarization) 
and on the atomic structure itself. 
In disordered structures and for small values of the wavevector 
magnitude ($ka\ll 1$), the spectral density  
both for longitudinal and transverse polarizations has  
the  shape of  a single pronounced peak 
(see Figs.~\ref{f17}(a) and \ref{f17}(d)). 
The positions of these low-frequency peaks at 
$\nu_{\rm t,l}$ are related to the wavevector magnitude 
according to the linear dispersion relation (see Fig.~\ref{f18}), 
\begin{equation}
\nu_{\rm t,l}\simeq c_{\rm t,l} k/2\pi
\ . 
\label{e4_3a}
\end{equation}
As seen from Fig.~\ref{f18}, the calculated dots in 
the low-frequency range lie on the straight lines 
plotted using experimental sound velocities  
($c_{\rm t} \simeq  37.5\AA/$ps   
and $c_{\rm l}\simeq 59\AA/$ps \cite{Benassi}). 

With an increase of the magnitude $k$ of  the wavevector, 
the peak-shaped spectral density shifts to higher 
frequencies and its width increases 
(see Figs.~\ref{f17}(b) and \ref{f17}(e)). 
At large enough $k\ge 1\AA^{-1}$, the spectral density no 
longer consists of   a 
single peak but rather resembles the vibrational 
density of states (VDOS)  
(see Figs.~\ref{f17}(c) and \ref{f17}(f)), 
clearly showing the two frequency bands found in 
v-SiO$_2$ \cite{PRB2}. 

If the spectral densities are peak-shaped, 
two of their characteristics, the peak position and  width, 
 are normally used  in order to describe  the 
propagation of external plane-wave excitations 
\cite{Allen,DelAnna,Benassi,Foret}.  
The peak position is associated with the average frequency of 
the propagating excitation,  while the peak width is associated 
 with the decay time 
of the excitation.  
Indeed, if we look at the ${\bf k}$-plane-wave component in the 
propagating excitation, ${\bf u}(t)$, 
its evolution with time is described by 
relation (\ref{e3_7}) at ${\bf k'}={\bf k}$. 
The weight (amplitude) of this component, 
$a_{\bf k}(t)$, decays with time according to 
Eqs.\ (\ref{e3_10}) and (\ref{e3_12}). 
A rough estimate of the time dependence of 
$a_{\bf k}(t)$ can be obtained if we assume that 
$a_{\bf k}(t) \propto a_{\bf k,{\rm c}}(t)$, 
i.e. $a_{\bf k}(t)$ is approximately the back cosine Fourier 
transformation of the spectral density 
$ {\overline\alpha}^j_{\bf k} 
{\underline\alpha}^j_{\bf k} $ (see Eq.~(\ref{e3_12})). 
If the spectral density has the shape of a 
 well-defined peak 
which can be fitted, say, by a Lorentzian, i.e. 
\begin{equation}
f_{\rm L}={1\over \pi}{\left( \Gamma_{\omega}/2 
\right) \over 
\left(\omega -\overline{\omega}_k\right)^2 + 
\left( \Gamma_{\omega}/2 \right)^2 } 
\ , 
\label{e4_4}
\end{equation}
where the Lorentzian position $\overline{\omega_k}$ 
and full-width at half-maximum (FWHM)  
$\Gamma_{\omega}$ are the fitting parameters, 
then the back cosine Fourier transform of the 
function (\ref{e4_4}) is 
\begin{equation}
a_k(t) \simeq A_2 \cos \overline{\omega}_k t 
\exp\left\{- \Gamma_{\omega}t/2 \right\} 
\ ,  
\label{e4_5}
\end{equation}
where we have actually used Eq.\  (\ref{e4_4}) to fit  the 
spectral density 
$|{\overline\alpha}^j_{\bf k}|^2$ 
normalized to unity  and then took into account the factor 
$A_2$ (see the beginning of the section). 
As clearly seen from Eq.~(\ref{e4_5}), the decay of the 
${\bf k}$-plane-wave component can be characterized by 
the average radial frequency 
$ \overline{\omega}_k$ and the inverse decay time 
\begin{equation}
\tau^{-1}_k \simeq \Gamma_{\omega}(k)/2 = \pi \Gamma_{\nu}(k)
\ ,  
\label{e4_6}
\end{equation}
with $ \Gamma_{\nu}$(THz)$=\Gamma_{\omega}/2\pi$.  

In Refs.~\cite{Benassi,DelAnna}, 
the damped harmonic  oscillator (DHO) model 
has been used to fit spectral densities, which gives similar 
values for the average frequency and width, if 
$(\Gamma_{\omega})^2 \ll \overline{\omega}_k^2 $. 
This inequality holds true in the region $k\le 1\AA^{-1}$ and 
in particular in the IR regime around $k\sim 0.1\AA^{-1}$ 
(see below),  
where the spectral densities 
have a well-defined peak shape and fitting of 
the spectral densities by 
 Lorentzian and/or DHO curves makes sense. 

We have used  fits both by the Lorentzian and DHO models to obtain 
the average frequency and decay time  (not shown, see 
Ref.~\cite{JPCM_99} for more detail) of the propagating 
plane-wave excitation 
 as a function of the initial wavevector. 
The dependence of $\overline{\nu}_k= \overline{\omega}_k/2\pi $
 vs. $k$ shown in Fig.~\ref{f18}
 can be associated with some sort of ''dispersion law``.  
Of course, the propagating plane-wave excitation cannot be 
characterized by only one wavevector (and single frequency) 
 and instead consists of a  
packet of plane waves (see Eq.~(\ref{e3_6})) with different 
wavevectors (packet of eigenmodes characterized by different 
frequencies). 
We chose from the ${\bf k'}$-packet only one component 
characterized by the 
same wavevector as that of the  initial plane wave and followed 
 its time evolution. 
In that case, the dependencies $\overline{\nu}_k$  
presented in  Fig.~\ref{f18} 
can be regarded as  the dispersion laws 
for a single plane-wave component. 
The experimental data  for longitudinal 
external plane-wave excitations from  IXS experiments 
\cite{Benassi}, 
obtained  by 
fitting the experimental curves with the DHO model, 
 are shown by the stars in 
Fig.~\ref{f18}(b) and they agree  well with our results. 

Note that the dispersion laws for  both branches 
 are practically linear in the 
 low-frequency (long-wavelength) regime for $\nu \le 3$THz. 
Above this frequency, a sort of ''fast-sound`` behaviour is observed. 
The increase in the slope of $\overline{\nu}_k$ 
is related to the changes in the 
shape of the spectral densities. 
A shoulder on the high-frequency side of the spectral-density peak 
for the longitudinal branch starts to appear at $k\ge 0.3\AA^{-1}$ 
($\nu \ge 0.3$THz - see Figs.~\ref{f17}(d),(e)). 
A similar transformation happens with the peak for 
the transverse branch 
at  $k\ge 0.5\AA^{-1}$. 

%
%
%
%
\subsection{Disordered zig-zag chain}
\label{S3_2}
\begin{figure}[b!] %
\centerline{\epsfig{file=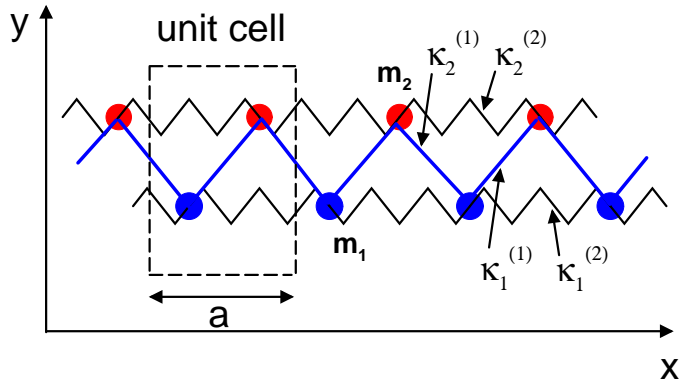,scale=0.5,angle=270}}
\vspace{1pt}
\caption{Zig-zag chain with two atoms in the unit cell. 
}
\label{f1}
\end{figure}

%
%
%
%
%
%
The other structural model we consider here is 
a toy model, namely a linear zig-zag chain on the plane. 
Toy models are very useful in studying atomic dynamics. 
Usually, scalar models are investigated because of their simplicity 
\cite{Schirmacher,Nakayama,Parshin_98}. 
However,  important effects related to  mixing of  modes of different 
polarizations \cite{JPCM_99} are missed in such models.  
That is why we consider below one of the simplest vectoral 
models, namely a zig-zag chain in the $x-y$ plane  (see Fig.~\ref{f1}). 
%
%
%
%

Consider a chain with 2 atoms in the unit cell: 
$ {\bf r}_1^{(0)} \{ x_1, y_1 \}$ and 
$ {\bf r}_2^{(0)} \{ x_2, y_2 \}$,  
so that the positions of  the other atoms can be found  as 
$ {\bf r}_{i,1}^{(0)} = {\bf r}_{1}^{(0)}  
+i {\bf a} $ and 
$ 
{\bf r}_{i,2}^{(0)} = {\bf r}_{2}^{(0)}  
+i {\bf a}
$ 
  with $ i=0, \pm 1, \dots   
$, 
where ${\bf a}=\{a, 0\}$ is a unit-cell vector, so that  
the chain is directed along the $x$ axis 
(see Fig.~\ref{f1}). 

The nearest neighbours of  different types (atoms 
$1$ and atoms $2$)  
are connected by  springs with  spring 
constants $\kappa^{(1)}_{1,i}$ in the same unit cell $i$ and with 
$\kappa^{(1)}_{2,i}$ in  different unit cells ($i$ and $i+1$), while 
 the nearest neighbours of the same 
type are linked 
by  springs with constants 
$\kappa^{(2)}_{1,i}$ for  atoms of the first type 
(the lower row) and 
$\kappa^{(2)}_{2,i}$ for the atoms of the second 
type (upper row). 

The potential energy then has the following 
expression: 
\begin{eqnarray}
V({\bf r}_{1,1}, {\bf r}_{1,2}, \dots )  
&=& 
{1 \over 2}\sum_i^{N_{\rm uc}} \Large\{ 
\kappa^{(1)}_{1,i}
\left( 
 |{\bf r}_{i,2}- {\bf r}_{i,1}| - 
|{\bf r}^{(0)}_{i,2}- {\bf r}^{(0)}_{i,1}|
\right)^2 
\nonumber  
\\ 
&+&  
\kappa^{(1)}_{2,i}
\left( 
 |{\bf r}_{i+1,1}- {\bf r}_{i,2}| - 
|{\bf r}^{(0)}_{i+1,1}- {\bf r}^{(0)}_{i,2}|
\right)^2   
\nonumber  
\\ 
&+&
\sum_{j=1,2}\kappa^{(2)}_{j,i}
\left( 
 |{\bf r}_{i+1,j}- {\bf r}_{i,j}| - 
|{\bf r}^{(0)}_{i+1,j}- {\bf r}^{(0)}_{i,j}|
\right)^2    
\Large\} \ , 
\label{e3_1_1}  
\end{eqnarray} 
with $i$ numbering all unit cells. 

%
%
%
%
\begin{figure}[b!] %
\centerline{\epsfig{file=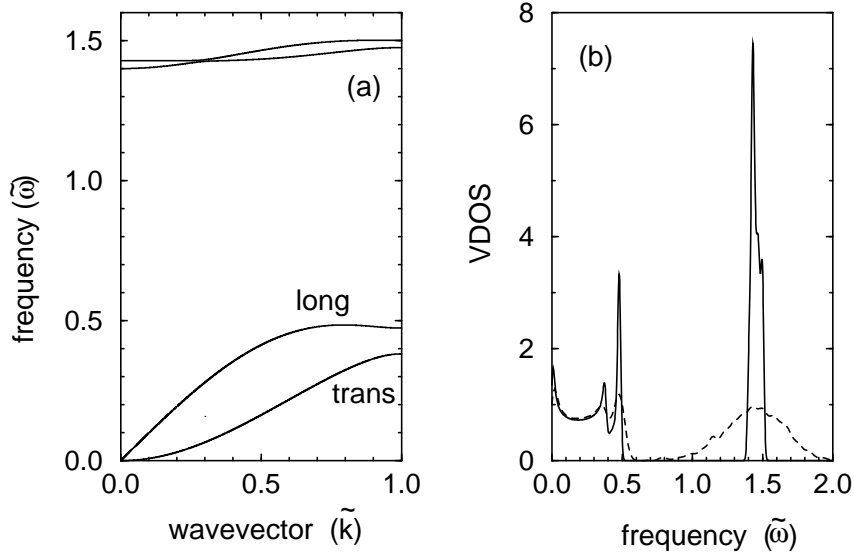,scale=0.5,angle=270}}
\vspace{10pt}
\caption{
(a): Dispersion curves 
($\tilde{\omega}=\omega \protect\sqrt{m/\kappa^{(1)}}$ vs 
$\tilde{k}=k/(\pi/a)$) for the zig-zag  
chain model  
 characterized by the following parameters: 
the equilibrium coordinate of atoms in the unit cell, 
${x_1/a=0; y_1/a=0}$, 
${x_2/a=0.4; y_2/a=0.5}$, ratio of force constants  
$\kappa^{(1)}/\kappa^{(2)}=0.1$ ($\kappa^{(1,2)}_1 =\kappa^{(1,2)}_2 $),  
 masses $m_1=m_2=m$ and the total number of atoms $N=2000$.   
(b): The VDOS of the  zig-zag chain model 
with the same set of parameters as in (a) (solid line) 
together with that for a disordered chain with fluctuations 
in force constants 
$\delta\kappa^{(1)}/\kappa^{(1)}=0.3$, 
$\delta\kappa^{(2)}/\kappa^{(2)}=0.3$ (dashed line).
}
\label{f3}
\end{figure}

Four branches occur  in the dispersion relation: 
two acoustic (longitudinal and transverse) and two 
optic branches (see Fig.~\ref{f3}a).  
The analytical expression for the dispersion is available 
for the symmetric case ($m_1=m_2=m; 
\kappa_{1,i}^{(1)} = \kappa_{2,i}^{(1)} = 
\kappa^{(1)}; 
\kappa_{1,i}^{(2)} = \kappa_{2,i}^{(2)} = 
\kappa^{(2)}; x_1/a=y_1/a=0; x_2/a=0.5$): 
\begin{equation}
{(\omega(k))^2 \over (\kappa^{(1)}/m) } = 
1 + {A\over 2} \pm \left[ 1+ {A^2\over 4} \pm 
A\cos{(ka/2)} \right]^{1/2} \ , 
\label{e3_1_2}
\end{equation}
with 
$
A = 2 
\left( 1 - \cos{ (ka)} \right)(\kappa^{(2)}/ \kappa^{(1)})$. 
From this expression it is not difficult to obtain 
that in the low-frequency limit 
($\omega \to 0$) for the longitudinal acoustic branch not surprisingly  
$\omega \propto k$, while for the 
transverse acoustic branch 
$\omega \propto k^2$. 
Such a $k^2$-dependence is typical for transverse 
vibrations of a linear chain and is related to the much 
smaller restoring force in the $y$-direction 
(compared to that in the $x$-direction) 
for long-wavelength vibrations because of the absence of  the 
spring continuum in that direction.   

The vibrational density of states $g(\omega)$ 
contains two bands characterized by typical van Hove singularities 
around the band boundaries (see Fig.~\ref{f3}b). 
The $k^2$-dependence of the transverse acoustic 
branch  results in an $\omega^{-1/2}$ singularity 
of the VDOS 
as  the frequency approaches zero, $\omega \to 0$. 

%
%
%
%
%
\begin{figure}[b!] %
\centerline{\epsfig{file=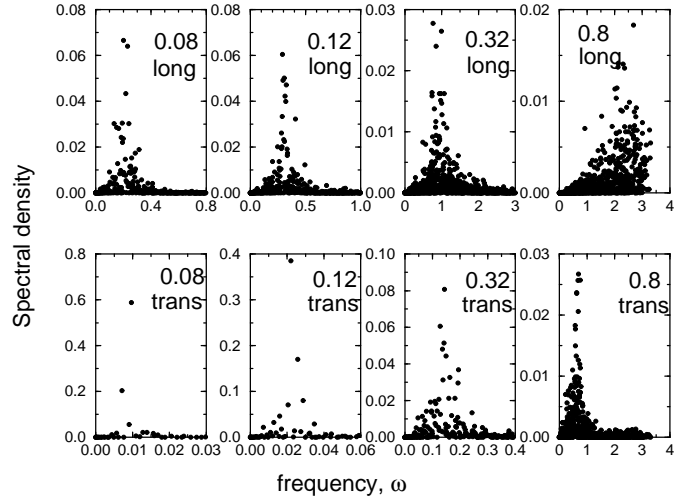,scale=0.4,angle=270}}
\vspace{10pt}
\caption{Spectral densities for a disordered zig-zag chain described 
by the following parameters: 
$m_1=m_2=1, \delta m_1=\delta m_2 = 0$,  
$
\kappa^{(1)}=\kappa^{(2)}=1, 
\delta\kappa^{(1)}=\delta\kappa^{(2)} = 1.0$, $ 
x_1/a=y_1/a=0, x_2/a = y_2/a = 0.5, a=1$ 
and $N=1000$. 
The wavevector magnitude, $k$, and polarizations 
(longitudinal or transverse) of the initial plane wave 
are marked in the figure. 
}
\label{f2}
\end{figure}

We are interested mainly in disordered structures. 
Several  different types of disorder can be 
introduced in the system:
 
(i) Positional disorder which is due to 
random positional vectors ${\bf r}_i^{(0)}$.  

(ii) Force-constant disorder: 
$\kappa_i$ are randomly distributed around their mean value, 
${\overline \kappa}$,  
\begin{equation}
\kappa_i = {\overline \kappa} + \delta\kappa\cdot\delta_{i,{\rm ran}} \ , 
\label{e3_1_3}
\end{equation}
where $\delta\kappa$ is a typical  width  
of the distribution and $\delta_{i,{\rm ran}}$ are   random numbers   
distributed around zero with the  density 
distribution, $\rho(\delta)$, e.g. the normal 
distribution, 
$
\rho(\delta)=\exp{\{-\delta^2/2\} }/ \sqrt{2\pi} \ , 
$
with  unit variance. 
Possible negative values of the spring constants are  
replaced by their absolute 
values.

(iii) Mass disorder: the atomic masses are randomly 
distributed, $m_i={\overline m}+\delta m 
\cdot \delta_{i,{\rm ran}}$. 

In the case of force-constant and mass disorder 
(which are considered 
in what follows), the equilibrium 
positions of the atoms are not changed. 
This is quite convenient and 
an ideal (crystalline) linear chain can be treated 
as the  crystalline 
counterpart of the disordered chain.  

The VDOS of a disordered zig-zag chain is 
similar to that for the crystalline chain 
(see Fig.~\ref{f3}b) except all sharp features 
(excluding the region around zero) are washed 
out (band tails appear). 
We should also note that  
extra states  (relative to the Debye spectrum)   
appear in the low-frequency regime ($\omega \le  
c_{\rm t,l}(\pi / a)$, with $c_{\rm t,l}$ 
being the transverse or longitudinal sound velocity. 
These extra states, characterized by the change in 
the VDOS, $ \Delta g(\omega) = g(\omega) - g_{\rm cryst} 
(\omega)\ ,  
$ 
are called the boson peak \cite{PRB2}. 
  
The  spectral densities for the disordered zig-zag chain, 
being of  
particular interest, are calculated according to 
Eq.~(\ref{e3_3}) and presented in Fig.~\ref{f2}. 

As  expected in the long-wavelength limit, 
the spectral density has the shape 
of a peak. 
The position of the peak can be approximately found 
from the dispersion for acoustic branches (see below). 
The width of the peak increases with  increasing magnitude  
of the wavevector of the initial plane wave. 
This corresponds to   more intense scattering 
of the plane waves by disorder. 
%
%
%
%
%
%
%
\section{ Analysis of the final state in  momentum space}
\label{S4}
As is well known, a scattering process can be investigated by 
analysis of the final state. 
First, we consider the final state of a single 
${\bf k}$-plane-wave component characterized by the same wavevector 
as the  initial one. 
The phase of this wave has a random value and is not an 
informative characteristic. 
The important quantity is the amplitude of the wave, or more 
precisely its squared average value, defined by 
Eq.~(\ref{e3_17}) with  
${\bf k'}={\bf k}$, 
\begin{equation}
\overline{a^2_{\bf k} }
= {1\over 2} \left\{   
{\sum_{j}|{\overline\alpha}^j_{\rm {\bf k}}|^2
|{\underline\alpha}^j_{{\rm {\bf k}},{\rm s}}|^2 
\over (\langle {\bf w}_{\rm {\bf k},s}^2 \rangle )^2 } 
+  
{\sum_{j}|{\overline\alpha}^j_{\rm {\bf k}}|^2
|{\underline\alpha}^j_{{\rm {\bf k}},{\rm c}}|^2 
\over (\langle {\bf w}_{\rm {\bf k},c}^2 \rangle )^2 }  
\right\} \simeq 
{\sum_{j}|{\overline\alpha}^j_{\bf k}|^2
|{\underline\alpha}^j_{\bf k}|^2 
\over  (\langle {\bf w}_{\bf k}^2 \rangle)^2 } 
\ .   
\label{e6_1}
\end{equation}
This value can be easily estimated for a peak-shaped spectral density 
of width $\Gamma$. 
Indeed, the number of eigenmodes contributing to an initial plane wave is 
of order  
$3N \cdot (\Gamma/D)$, where $D$ is the width of the whole 
vibrational spectrum 
($\simeq 40$THz in the  case of vitreous silica). 
Then we can easily evaluate from the normalization conditions 
Eqs.\ (\ref{e3_5}), (\ref{e3_16})
for the spectral densities  the average value of 
the spectral density in the peak region, 
$ |{\overline\alpha}^j_{\rm {\bf k}}|^2 \sim 
|{\underline\alpha}^j_{{\rm {\bf k}},{\rm s}}|^2  
\sim (D/\Gamma)\cdot(1/3N)$,  
and obtain the following estimate 
for $\overline{a^2_{\bf k} }$, 
\begin{equation}
\overline{a^2_{\bf k} } \sim {D\over \Gamma}\cdot {1\over 3N} 
\ ,   
\label{e6_2}
\end{equation}
where we have taken into account that 
$ \langle {\bf w}_{\bf k}^2 \rangle \sim 1$ according to Eq.~(\ref{e3_5}). 
The factor $D/\Gamma$ in  relation\ (\ref{e6_2}) shows that 
the averaged squared amplitude is inversely proportional to the 
 number 
of  initially excited modes and not to all the  modes. 
The factor $D/\Gamma$ in the Ioffe-Regel (IR) region is much larger than unity;  
$D/\Gamma \sim 10^2$ for 
$\Gamma\sim \nu_{\rm IR}/\pi \sim 0.3$THz. 
Therefore we can say that the ${\bf k}$-plane-wave component 
around and below the IR regime is not damped (the squared average 
amplitude is not of order  $1/3N$) but rather is attenuated (scattered), 
weakly (strongly)  below (beyond) the IR limit as discussed below. 
The ${\bf k}$-plane-wave component is damped at $k\ge k_* \simeq 
1\AA^{-1}$ when the peak width becomes comparable  to the full spectral  
width, $\Gamma \sim D$. 

The analysis of all ${\bf k'}$-plane-wave components in the final state, 
in particular the distribution  (\ref{e3_17}) of their weights,  
allows  the scattering mechanism to be clarified.  
Let us consider an initial plane wave characterized by 
the wavevector ${\bf k}$ and polarization 
${\bf{\hat n}}$. 
This wave is scattered with time into different plane 
waves characterized by wavevectors 
${\bf k'}$ and polarizations ${\bf{\hat n}'}$, which 
do not necessarily coincide with the initial polarization. 
We would like to know the weights of all 
plane-wave components in the final state as a function of the 
wavevector magnitude $k'$.   
The distributions of the transverse and longitudinal  plane waves, 
 $\rho({\bf k'},{\bf{\hat n}'}_{\rm t}|{\bf k},{\bf {\hat n}})$ and 
$\rho({\bf k'},{\bf{\hat n}'}_{\rm l}|{\bf k},{\bf {\hat n}})$ 
(see Eq.~(\ref{e3_18})), 
and the distribution 
averaged over  polarizations, 
 $\rho_{\rm av}({\bf k'}|{\bf k},{\bf {\hat n}})$ 
(see Eq.~(\ref{e3_19})),  for 
both transverse ${\bf {\hat n}}_{\rm t}$ and longitudinal 
 ${\bf {\hat n}}_{\rm l}$ polarizations of the initial plane-wave 
excitation  are of  particular interest. 
These distributions depend only on the spectral 
densities 
$ |{\overline\alpha}^j_{\bf k}|^2 $, 
$ |{\underline\alpha}^j_{\bf k'}|^2 $ and the vibrational 
spectrum itself,  and 
can be easily calculated numerically for different $k$. 
%
%
%
%
%
%
\subsection{Vitreous silica} 
\label{S4_1}
%
%
%
%
%
%

\begin{figure}[b!] %
\centerline{\epsfig{file=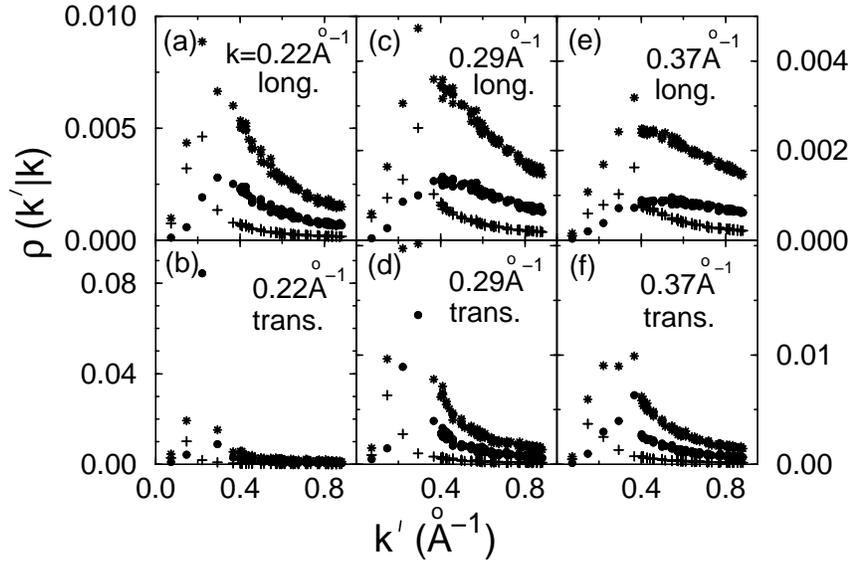,scale=0.5,angle=270}}
\vspace{10pt}
\caption{
The distribution functions  
$\rho({\bf k'},{\bf{\hat n}'}_{\rm t}|{\bf k},{\bf {\hat n}})$ (circles),  
$\rho({\bf k'},{\bf{\hat n}'}_{\rm l}|{\bf k},{\bf {\hat n}})$ (pluses) and 
$\rho_{\rm av}({\bf k'}|{\bf k},{\bf {\hat n}})$ (stars) 
for longitudinal ((a), (c) and (e)) and transverse 
((b), (d) and (f)) initial 
polarizations of plane waves characterized 
by different initial wavevector magnitudes 
$k$ for a  structural model of v-SiO$_2$. 
}
\label{f23}
\end{figure}

The results of such calculations for vitreous silica 
 are  presented  
 in Fig.~\ref{f23}. 
The upper (lower) row describes the scattering of initial 
longitudinal (transverse) plane waves, characterized by different 
wavevector magnitudes, into transverse 
and longitudinal   plane waves and also the distribution of the 
weights averaged over the 
polarization in the final state. 

%
%
%
%

First, we consider scattering of a longitudinal initial wave 
(the upper row in  Fig.~\ref{f23}). 
The weight distributions, 
 $\rho({\bf k'},{\bf{\hat n}'}_{\rm l}|{\bf k},{\bf {\hat n}}_{\rm l})$
and 
$\rho({\bf k'},{\bf{\hat n}'}_{\rm t}|{\bf k},{\bf {\hat n}}_{\rm l})$, 
characterize 
the scattering of the longitudinal wave to a longitudinal wave, 
the $\{ l \to l \}$ channel, 
and of the longitudinal wave to a transverse wave, the  $\{ l \to t \}$ 
channel, respectively. 
As follows from Fig.~\ref{f23}, these distributions are peak shaped but 
the positions of the peaks are different. 
The distribution for the $\{ l \to l \}$ channel  has a maximum around 
$k'_{\rm ll} \simeq k_{\rm l}\equiv k$ (or maybe a bit below the initial wavevector), 
while the  distribution for the $\{ l \to t \}$ 
channel   is mainly concentrated at a higher wavevector 
value, $k'_{\rm lt} > k_{\rm l}$. 
The distribution, $\rho_{\rm av}({\bf k'}|{\bf k},{\bf {\hat n}}_{\rm l})$, 
averaged over polarizations  of the final state 
is a sum of double the  distribution for the $\{ l \to t \}$ channel plus  
the distribution for the $\{ l \to l \}$ channel. 
If the peaks related to the individual channels and 
constituting the average distribution are narrow enough, then the 
distribution function  $\rho_{\rm av}({\bf k'}|{\bf k},{\bf {\hat n}}_{\rm l})$ 
    is doubly peaked (not clearly seen in Fig.~\ref{f23}). 
If the peaks  are too wide, then  $\rho_{\rm av}({\bf k'}|{\bf k},{\bf {\hat n}}_{\rm l})$ 
looks like a single wide peak (see Fig.~\ref{f23}) with 
a maximum position $k'_{\rm l,av}$ close to 
$k'_{\rm lt}$. 

Such a shape of the distributions of the weights of plane 
waves in the final state can be qualitatively understood 
in  the following way. 
The distribution function 
$\rho({\bf k'},{\bf{\hat n}'}_{\rm t}|{\bf k},{\bf {\hat n}}_{\rm l})$ 
of the transverse waves is an integral 
(sum in the case of a finite-size model)  of the product 
of two spectral densities, 
$ |{\overline\alpha}^j_{{\bf k},{\bf{\hat n}}_{\rm l}}|^2 $  
for longitudinal and  
$ |{\underline\alpha}^j_{{\bf k'},{\bf{\hat n}'}_{\rm t}}|^2 $   
for transverse polarization. 
Around the IR region and below it, 
these peak-shaped spectral densities have  maxima 
at $\nu_{\rm l} \simeq c_{\rm l}k/2\pi$ and 
$\nu'_{\rm t} \simeq c_{\rm t}k'/2\pi$, respectively, 
which generally do not coincide with each other. 
Therefore, the distribution 
$\rho({\bf k'},{\bf{\hat n}'}_{\rm t}|{\bf k},{\bf{\hat  n}}_{\rm l})$ 
has a maximum around $k'_{\rm lt}$ satisfying the 
equation, 
$ \nu_{\rm l}\simeq c_{\rm l}k/2\pi  \simeq  
c_{\rm t}k'_{\rm lt} /2\pi \simeq \nu'_{\rm t}$, i.e.    
\begin{equation} 
k'_{\rm lt} \simeq c_{\rm l}k/c_{\rm t} 
\ , 
\label{e6_3}
\end{equation}
which is obviously greater than  the wavevector 
of  the initial longitudinal wave. 

The distribution of longitudinal waves for the  $\{l\to l\}$ 
channel can be analysed in a similar manner. 
The main difference from the $\{l\to t\}$ channel  is that 
the spectral density  of the  longitudinal plane wave in the 
final state coincides with the spectral density of the 
initial longitudinal plane wave at approximately  the same 
wavevector magnitude as for the initial wave, 
\begin{equation} 
k'_{\rm ll}\simeq k
\ .  
\label{e6_4}
\end{equation}
Actually, the value $ k'_{\rm ll}$ should be slightly 
shifted to lower values, because the height of the peak for the 
spectral density 
$ |{\underline\alpha}^j_{\bf k'}|^2 $ 
increases  with decreasing  $k'$ and the maximum 
of the product of the spectral densities is reached 
in the low-frequency tail of the spectral density for the initial 
plane wave. 

The scattering of an initially transverse plane wave occurs 
similarly. 
In particular, the conclusion that the average frequency, $\nu'$, 
 of the 
majority of the plane-wave components comprising the final state coincides with 
the average frequency, $\nu$,  of the initial plane wave, 
\begin{equation} 
\nu' \simeq \nu 
\ ,   
\label{e6_5}
\end{equation}
holds true independently of the polarization of the 
initial plane-wave excitation. 
Therefore, we can roughly say that the disorder-induced scattering of the 
plane wave is approximately ''elastic`` (on average). 
This is not an absolutely precise conclusion because, first, the plane-wave 
components are distributed in frequency 
(composed of  eigenmodes having different frequencies) 
 in the initial and final states and, 
second, 
even the maximum of the distribution in the final state 
 is slightly shifted to lower frequencies 
as compared to the initial one, as discussed above. 

In the case of the scattering of an initial  transverse plane wave, 
two channels are available: $\{t\to l\}$ and 
$\{t\to t\}$. 
The distribution functions, 
$\rho({\bf k'},{\bf {\hat n}'}_{\rm l}|{\bf k},{\bf{\hat  n}}_{\rm t})$ 
and 
$\rho({\bf k'},{\bf{\hat n}'}_{\rm t}|{\bf k},{\bf{\hat n}}_{\rm t})$, 
of the weights of plane waves in the final state for these 
channels have peaks located around the following 
values: 
\begin{equation} 
k'_{\rm tl} \simeq c_{\rm t}k/c_{\rm l} \ \ \ 
\mbox{and} \ \ \ 
k'_{\rm tt} \simeq k
\ .  
\label{e6_6}
\end{equation}
As follows from Eq.~(\ref{e6_6}) and Figs.~\ref{f23}(b),(d),(f),  
the peak for longitudinal waves lies below the initial $k$,  
while for transverse waves the peak approximately 
coincides with $k$, being slightly shifted to smaller 
values for   reasons  similar to those  discussed above 
for the $\{l\to l\}$ channel. 

The distribution functions shown in Fig.~\ref{f23} 
were obtained for  a bar-shaped structural model of 
v-SiO$_2$. 
Such a model is effectively one-dimensional and 
has  restrictions  for the available initial ${\bf k}$ 
and final ${\bf k'}$ vectors,  
which are mainly directed along the bar in the low-$k$ limit. 
This also restricts the number of   scattering 
channels. 
In order to check the influence of the dimensionality 
of the model on the scattering 
of plane waves, we have performed a similar 
analysis for a cubic ($3$-dimensional) model of v-SiO$_2$ and 
 have not found 
any influence of the dimensionality of the model 
for the available  wavevector magnitudes $k\ge 0.22\AA^{-1}$ 
(for the cubic model).  

Poor statistics in the long-wavelength limit 
(see Fig.~\ref{f23}) is a finite-size effect 
related to the restricted number of the wave-vectors  
allowed by the periodic boundary conditions. 
An analytical extrapolation approach \cite{JPCM_99} can be used 
to overcome such a shortcoming. 

%
%
%
%
%
%
\subsection{Zig-zag chain} 
\label{S4_2}
%
%
%
%
%
%
%

Another possible way to overcome the disadvantages of  
finite-size $3$-D numerical models is  to  analyse 
low-dimensional models. 
Much lower wavevectors $k\ge k_{\rm min}^{(d)} = 2\pi/N^{1/d}a$ 
 are available, for example,   
in one-dimensional ($d=1$) models as compared to the 
$3$-D case, and the acoustic spectrum appears to be 
be much more dense.  
In order to check and support the analytical and numerical 
 approaches presented above for the 
$3$-D case,  we have performed  numerical 
experiments for a disordered $1$-D  zig-zag chain (see Sec.~\ref{S3_2})
 and calculated 
the distribution function 
$ \rho({\bf k'},{\bf{\hat n}'}|{\bf k},{\bf{\hat n}})$ for it. 

%
%
%
%
\begin{figure}[b!] %
\centerline{\epsfig{file=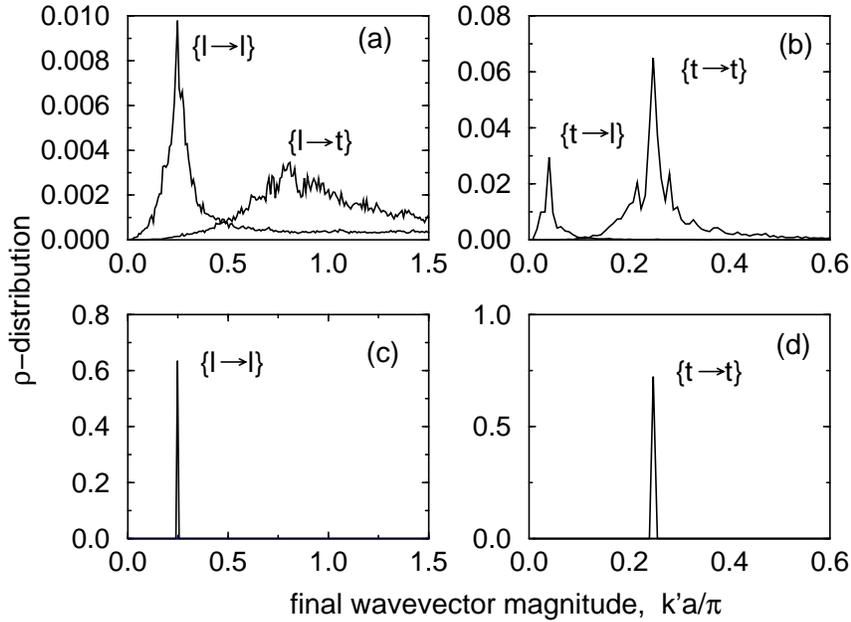,
scale=0.5,angle=270}}
\vspace{10pt}
\caption{The distribution function 
$ \rho({\bf k'},{\bf{\hat n}'}|{\bf k},{\bf{\hat n}})$ 
for different scattering channels (as marked in the figure) 
for a 
disordered ((a) and (b)), $\delta\kappa^{(1)} = 
\delta\kappa^{(2)} = 1$, $\delta m_1 = \delta m_2 =0$) 
and an ordered ((c) and (d)) zig-zag 
chain characterized by the following parameters: 
$m_1=m_2=1$, $\kappa^{(1)}=\kappa^{(2)}=1$, $x_1/a=y_1/a=0$, 
$x_2/a=y_2/a=0.5$. 
The initial wavevector magnitude $ka/\pi 
= 0.25$. 
}
\label{f25}
\end{figure}
%
%

%
%
%
%
\begin{figure}[b!] %
\centerline{\epsfig{file=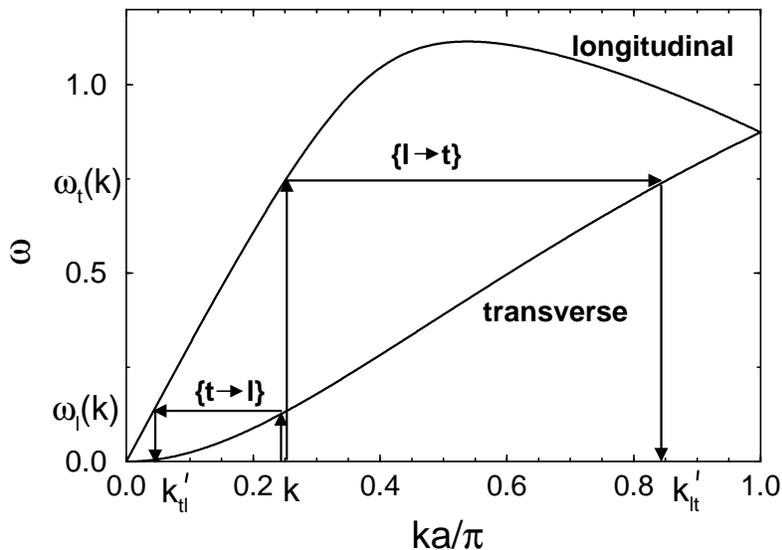,
scale=0.5,angle=270}}
\vspace{1pt}
\caption{ The acoustic branches for an ordered zig-zag chain characterized 
by the same parameters as in Fig.~\protect\ref{f25}. 
The construction used to estimate values of $k'_{\rm tl}$ and $k'_{\rm lt}$ 
is shown. 
}
\label{f24}
\end{figure}

Our main purpose  here   is to calculate 
the distribution function 
$ \rho({\bf k'},{\bf{\hat n}'}|{\bf k},{\bf{\hat n}})$ characterizing 
the scattering of  a plane-wave excitation. 
First, we have calculated this distribution function 
for the crystalline counterpart ($\delta\kappa_i=0$) 
and not surprisingly  we 
 found for $k\le \pi/a$ only $\{ t \to t \}$ and 
$\{ l \to l\}$ channels (see the lower row in  
Fig.~\ref{f25}  marking the peaks at 
$k'_{\rm tt}=k$ and $k'_{\rm ll }=k$ for the 
$\{ t \to t \}$ and $\{ l \to l\}$ channels, respectively).  
Disorder changes the situation dramatically and gives rise to 
the occurrence  of  
 $\{ t \to l \}$ and $\{ l \to t\}$ channels 
(see the upper row in Fig.~\ref{f25}), in complete 
agreement with the results of the $k$-analysis 
given  
in Fig.~\ref{f23} for the case of v-SiO$_2$. 
The positions of  the additional peaks, at 
$k'_{\rm tl}$ ($\{ t \to l \}$ channel), and 
$k'_{\rm lt }$ ($\{ l \to t \}$ channel) can be obtained  
from the  dispersion laws for the crystalline zig-zag chain by solving 
the equations: 
\begin{equation}
\omega_{\rm t}(k) = \omega_{\rm l}(k'_{\rm tl}) \ ,
\label{1}
\end{equation}
and 
\begin{equation}
\omega_{\rm l}(k) = \omega_{\rm t}(k'_{\rm lt}) \ , 
\label{2}
\end{equation}
respectively (see arrows in Fig.~\ref{f24}). 
The width of the peaks   increases  with  
increasing  disorder. 
We have also found a similar shape of the distribution 
function $\rho$ (for four channels)  for all wavevectors 
$k\le \pi/a$ with the corresponding $\omega_{\rm t}$ and 
$\omega_{\rm l}$ lying in the range of the dense spectrum.  
%
%
%
%
%
%
%
\section{Scattering mechanism} 
\label{S5}
%
%
%
%
%
%
%
From numerical calculations for both  
the  $1$-D zig-zag chain model and the $3$-D model of v-SiO$_2$, 
we have found that plane waves  
scatter not only to  modes of approximately 
the same wavelength (and polarization) but also to  modes 
of  different wavelength (and polarization) but of similar  frequency. 
The reason for such scattering is a natural question. 

In our simulations on zig-zag chains, we have 
found the $\{t\to l\}$ and 
$\{l\to t\}$ scattering channels even for models 
not showing an appreciable increase of the VDOS 
in the low-frequency regime. 
Hence, the  appearance  of the $\{t\to l\}$ and 
$\{l\to t\}$ channels should be explained in terms  
of existing  transverse and longitudinal acoustic 
waves.  
Indeed, in the crystal, transverse and longitudinal  
acoustic phonons are orthogonal 
to each other, and hence transverse (longitudinal) plane 
waves cannot be scattered into longitudinal (transverse) plane waves 
(as we checked numerically; see the lower row in Fig.~\ref{f25}). 
Disorder leads to   changes in the interaction energy, 
with the result that an acoustic phonon with a particular energy 
can couple to other phonons with closely comparable energies, including 
 phonons with different polarizations and wavevectors. 
Therefore, the resulting eigenmodes contain components of different 
polarization 
and different wavevectors. 
A plane wave is not an  eigenmode in the disordered stucture and it 
is scattered into  eigenmodes of  approximately 
the same energy as  that of the plane wave. 
These eigenmodes contain both transverse and 
longitudinal components and therefore an original 
plane wave, independent of its polarization, is scattered into 
both transverese and longitudinal plane waves. 
This gives a qualitative explanation of the existence of  
$\{t \to l\}$ and $\{ l \to t \}$ scattering channels. 

In the $3$-D case, the situation can be more complicated. 
Apart from the scattering mechanism due to 
the disorder-induced mixing of transverse and 
longitudinal plane waves discussed above, 
  extra states  (comprising the Boson peak) relative to the Debye 
spectrum (e.g. optic modes pushed down by disorder 
(see \cite{JPCM_99,PRB_anh}))  
 could participate in the hybridization 
between plane waves with different polarization.


\begin{references} 

\bibitem{PingSheng_book1} {\it Scattering and Localization of Classical 
Waves in Random Media}, edited by   Sheng P. (World Scientific, Singapore, 1990). 

\bibitem{PingSheng_book2}  Sheng P., 
{\it Introduction to Wave Scattering, Localization, and Mesoscopic Phenomena} 
 (Academic Press, San Diego, 1995). 

\bibitem{Mott_Davis_book} Mott N.F. and Davis E.A., {\it Electronic Processes in 
Non-Crystalline Materials  }, $2$nd ed.  
(Clarendon Press, Oxford, 1979).

\bibitem{John}  John S.,  Sompolinsky H., and Stephen M.J., 
Phys.Rev. B {\bf 27}, 5592 (1983).

\bibitem{Aharony}  Aharony A.,  Alexander S., Entin-Wohlman O., 
and  Orbach R., Phys. Rev. Lett. {\bf 58}, 132 (1987).

\bibitem{Alexander}  Alexander S., Phys.Rev. B {\bf 40}, 7953 (1989). 

\bibitem{IR}  Ioffe A.F. and   Regel A.R., Prog. Semicond. {\bf 4}, 237 (1960). 

\bibitem{PingSheng_1994}  Sheng P., Zhou M., and Zhang Z.-Q.,  
Phys. Rev. Lett. {\bf 72}, 234 (1994). 

\bibitem{Anglaret} Anglaret E., Hasmy A., Courtens E., Pelous J., and Vacher R., 
Europhys. Lett. {\bf 28}, 591 (1994).  

\bibitem{Grest} Grest G.S., Nagel S.R., and Rahman A., Phys. Rev. Lett. 
{\bf 49}, 1271 (1982). 

\bibitem{Hafner} Hafner J. and Crajci M., J.Phys.: Cond.Matter,{\bf 6}, 4631 (1994).  

\bibitem{Schober_Liard} Schober H.R. and Laird B.B., 
Phys. Rev. B {\bf 44}, 6746 (1991). 

\bibitem{Gurevich_93} Gurevich V.L., Parshin D.A, 
 Pelous J., and Schober H.R., Phys.Rev., 
             B {\bf 48}, 16318 (1993).

\bibitem{Mazzacurati_EPL}  Mazzacurati V., Ruocco G., and 
 Sampoli M., Europhys.Lett., {\bf 34}, 681 (1996). 

\bibitem{Sampoli_98} Sampoli M., Benassi P., Dell'Anna R.,  
Mazzacurati V. and Ruocco G., Phil.Mag., B {\bf 77}, 473 (1998).

\bibitem{EPL} Taraskin S.N. and Elliott S.R., Europhys. Lett., {\bf 39}, 37 (1997). 

\bibitem{Madden} Ribeiro M.C.C., Wilson M., and Madden P.A., J.Chem.Phys., 
{\bf 108}, 9027 (1998). 

\bibitem{Allen1} Allen P.B. and Feldman J.L.,
Phys. Rev. B {\bf 48}, 12581 (1993),  

\bibitem{Allen} Feldman J.L., Kluge M.D., Allen P.B., and 
 Wooten F., Phys. Rev. B {\bf 48}, 12589 (1993).   

\bibitem{Fabian} Fabian J. and Allen P.B., 
Phys. Rev. Lett., {\bf 77}, 3839 (1996).  

\bibitem{DelAnna} Dell'Anna R., Ruocco G., Sampoli M., and Viliani G., 
Phys. Rev. Lett. {\bf 80}, 1236 (1998). 

\bibitem{Feldman_1998} Feldman J.L., 
 Allen P.B., and Bickham S.R., 
Phys. Rev. B {\bf 59}, 3551  (1999).  

\bibitem{Wischnewski_98} Wischnewski A., Buchenau U., Dianoux A.J.,
  Kamitakahara W.A., and Zarestky J.L., 
Phys. Rev.  B  {\bf 57}, 2663 (1998). 

\bibitem{PingSheng_1991}Sheng P. and Zhou M., 
Science {\bf 253}, 539 (1991). 

\bibitem{SM_phonons} 
Buchenau U., Galperin Yu.M., Gurevich V.L., Parshin D.A., 
 Ramos M.A. and Schober H.R., Phys.Rev., B {\bf 46}, 
2798 (1992). 

\bibitem{Burin} Gaganidze E., Konig R., Esquinazi P., Zimmer K., and Burin A., 
 Phys. Rev. Lett. {\bf 79}, 5038 (1997). 

\bibitem{Vacher} Vacher R., Pelous J., 
and Courtens E., Phys. Rev. B {\bf 56}, R481 (1997). 

\bibitem{Kob_JNCS} Horbach J., Kob W. and Binder K., J.Non-Cryst.Sol., {\bf 235}, 
320 (1998).   

\bibitem{Maradudin} Maradudin A.A., Montroll E.W., Weiss G.H., 
and Ipatova I.P., {\it Theory of Lattice Dynamics in the 
Harmonic Approximation} (Acad. Press, N.Y., 1971).

\bibitem{part1} Leibfried G. and Breuer N., {\it Point Defects in Metals I. 
Introduction to the theory} (Springer, Berlin, 1978). 

\bibitem{part2} Dederichs P.H. and Zeller R., in {\it Point Defects 
in Metals II. Dynamical Properties and Diffusion Controlled Reactions} 
(Springer, Berlin, 1980). 

\bibitem{Klinger_review} Klinger M.I., Phys.Rep., {\bf 165}, 275 (1988).

\bibitem{Karpov_review} Galperin Yu.M., Karpov V.G., and 
 Kozub V.I., Adv.Phys. {\bf 38}, 669 (1989). 

\bibitem{Schirmacher} Schirmacher W., Diezemann G., and Ganter C., 
Phys. Rev. Lett. {\bf 81}, 136 (1998).

\bibitem{SRE_book} Elliott S.R., {\it Physics of Amorphous Materials} 
$2^{\rm nd}$ Edn. 
(Longman, N.Y., 1990). 

\bibitem{Hansen_book} Hansen J.-P. and McDonald I.R., {\it Theory of Simple Liquids} 
$2$-nd Ed. (Academic Press, London, 1990). 

\bibitem{VB} van Beest B.W.H., Kramer G.J., and van Santen R.A.,  
Phys.Rev.Lett., 
{\bf 64}, 1955 (1990). 

\bibitem{Guissani} Guissani Y. and Guillot B., 
J.Chem.Phys., {\bf 104}, 7633 (1995).  

\bibitem{Kob} Vollmayr K., Kob W., and Binder K., Phys.Rev., 
              {\bf B54}, 15808 (1996).

\bibitem{PRB2} Taraskin  S.N. and  Elliott S.R., Phys.Rev., 
B{\bf 56}, 8605  (1997). 

\bibitem{Bermejo} Fernandez-Perera R., Bermejo F.J., and Enciso E.,  
Phys. Rev., B {\bf  53}, 6215 (1996).  

\bibitem{Benassi} Benassi P., Krisch M., Masciovecchio C., Mazzacurati V., 
 Monaco G., Ruocco G., Sette F., and Verbeni R., 
Phys. Rev. Lett. {\bf 77}, 3835 (1996). 

\bibitem{Foret}  Foret M., Courtens E., Vacher R., and 
 Suck J.-B., Phys. Rev. Lett. {\bf 77}, 3831 (1996). 

\bibitem{Nakayama} Nakayama T., Phys. Rev. Lett. {\bf 80}, 1244 (1998).

\bibitem{Parshin_98} Parshin D.A. and 
Schober H.R., Phys. Rev., B {\bf  57}, 10232 (1998).

\bibitem{JPCM_99}  Taraskin S.N. and  Elliott S.R., 
J.Phys.: Cond.Matter, {\bf 11}, A219 (1999). 

\bibitem{PRB_anh} Taraskin S.N. and Elliott S.R., Phys.Rev., 
B, {\bf  59}, 8572 (1999) .  


\end{references}
\end{document}